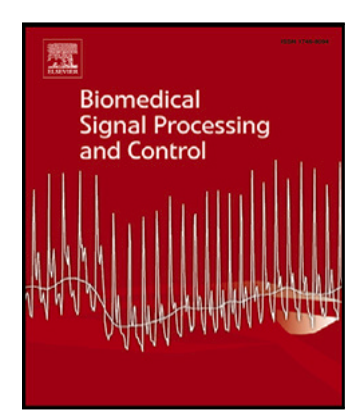

# Practical flow state detection: Entropy-based EEG classification from portable EEG headbands

Matin Beiramvand *, Reijo Koivula, Tarmo Lipping

*Faculty of Information Technology and Communication Sciences, Tampere University, Finland*

ABSTRACT

Flow state, characterized by deep engagement and immersion during challenging activities, represents a valuable mental state with significant implications for learning, performance, and rehabilitation outcomes. While flow has been extensively studied behaviorally, objective neurophysiological detection methods suitable for real-world deployment remain limited. Electroencephalography (EEG) offers a promising avenue for flow detection due to its accessibility, portability, and superior temporal resolution; however, the utility of consumer-grade EEG devices for robust, and subject independent flow classification has been insufficiently explored. This study validates entropy-based biomarkers for flow state detection using two wearable EEG headsets, Muse-S and Emotiv Insight, across 45 participants performing adaptive Tetris gameplay. After denoising, we applied the Discrete Wavelet Transform (DWT) to decompose the signals into multiple frequency sub-bands. From each sub-band, entropy-based features, combining channel-wise measures (Slope Entropy, Distribution Entropy, Spectral Entropy) with cross-channel descriptors (Cross Distribution Entropy, Cross Spectral Entropy) were extracted. These features were then used as input to a Random Forest classifier (RF), evaluated with two validation schemes: Random Sampling (RS) and Leave-One-Subject-Out (LOSO). Under random sampling cross-validation, Random Forest classifiers achieved 97% mean accuracy; under the more rigorous leave-one-subject-out (LOSO) scheme, average accuracy reached 75%, demonstrating genuine cross-subject generalizability. Comprehensive multi-classifier validation (SVM-RBF, GentleBoost, k-NN, Fitted Discriminant, Naive Bayes) confirmed that entropy biomarkers are robust across diverse modeling frameworks. These findings establish features as reliable, device-independent neural signatures of flow and demonstrate the feasibility of consumer EEG for practical flow detection in real-world scenarios, a critical advancement toward deployable neuroergonomic assessment, adaptive training systems, and personalized rehabilitation interventions.

## 1. Introduction

The theory of flow state was first introduced by Mihály Csíkszentmihályi in 1975, describing flow as a state characterized by automaticity, intrinsic reward, optimal performance, intense engagement, and a sense of control over oneself and the outcome of the activity [1]. Flow has been examined across a broad spectrum of activities including, ocean cruising [2], motorcycling [3], and various sports [4], as well as computer gaming [5,6], music performance [7], and literary writing [8]. Despite engaging different cognitive, emotional, and sensorimotor processes, the subjective experience of flow remains remarkably similar across these activities [9]. Physiological and neurophysiological biomarkers provide objective and real time indicators of the flow state. A range of such biomarkers, extracted from cardiac measures such as heart rate variability (HRV) and blood pressure (BP), respiratory parameters such as respiratory rate (RR), ocular metrics including blink rate (BR) and pupil diameter, and skin related variables such as skin temperature and electrodermal activity, have been the focus of extensive empirical investigation [10]. Flow during Tetris gameplay was examined while recording cardiac and respiratory activity, including ECG, HRV, and chest and abdominal respiration. Higher flow was linked to greater respiratory depth. It was also associated with reduced low-frequency power in HRV. The optimal difficulty condition yielded the highest state flow, positive affect, and effortless attention [11].

Flow versus non-flow was classified from cardiac features specifically HRV using a Random Forest (RF) [9]. Participants were divided into two cohorts: a field study, in which individuals were repeatedly surveyed while working in their natural environment, and a laboratory study, in which flow was elicited with an invoice matching task calibrated to balanced difficulty. Overall, the RF distinguished high from low flow, with higher accuracy in the field than in the lab [9]. In a study of 22 elite pianists, decreased LF power and increased HF power were observed before peak performances [12]. Although HRV is the most common ECG-derived measure in flow research, recent research has employed more sophisticated ECG analyses. For instance, during flow, heart-evoked potential (HEP) amplitude tends to decrease an indicator of reduced self-referential processing and HEP amplitude has been correlated with absorption in flow [13]. ECG analyses also suggest that experiencing flow may relate to depression and possibly anxiety, with neuroticism and familial factors acting as important confounders in observed associations between flow proneness and health outcomes [14].

* Corresponding author.
*E-mail address:* matin.beiramvand@tuni.fi (M. Beiramvand).

Cardiovascular-based measures show promise for assessing flow, however, combining them with other physiological signals has produced more reliable flow-state detection. The researchers used multiple wearable sensors to simultaneously record four-channel EEG, heart rate (HR) and blood oxygen saturation ($SpO_2$) via photoplethysmography (PPG), galvanic skin response (GSR), and motion tracking. Combining EEG, cardiovascular, electrodermal, and kinematic signals provided complementary information about flow-state characteristics [15].

Researchers have recently begun to investigate the neural correlates of this intriguing phenomenon, which has been the subject of numerous behavioral studies. Neurologically, flow has been linked to transient hypofrontality, a phenomenon characterized by a temporary decrease in activity within prefrontal brain regions [11]. This reduction dampens the analytical and self-reflective operations of the brain's explicit processing system while enhancing the use of procedural and skill-based knowledge managed by the implicit system [16]. Key brain regions implicated in attention regulation, executive functioning, and reward processing namely the dorsolateral prefrontal cortex (DLPFC), medial prefrontal cortex (MPFC), and inferior frontal gyrus (IFG) are critically involved in facilitating flow experiences [17].

Although several biomedical signals show promise for detecting flow, the phenomenon originates in the brain; most peripheral biosignals primarily index autonomic responses to being in flow. Functional magnetic resonance imaging (fMRI) and conventional MRI have yielded candidate neural markers, For instance, fMRI studies have shown that flow is associated with increased activation in the anterior insula, inferior frontal gyri, basal ganglia, and midbrain (relative to boredom and overload), and decreased activation in the medial prefrontal cortex, posterior cingulate cortex, and medial temporal lobes (including the amygdala) [18]. Yet, the evidence remains limited and inconsistent due to methodological constraints and variability across studies [19], and [11]. Accordingly, much of the literature emphasizes neurophysiological biomarkers, particularly electroencephalography (EEG) and functional near-infrared spectroscopy (fNIRS). Among neurophysiological modalities, EEG is the most accessible while providing excellent temporal resolution.

The researchers used multichannel EEG to monitor flow during tasks spanning gaming, creative challenges, and learning [20]. By analyzing dynamic power changes across frequency sub-bands and applying KNN, elastic net, random forest, and linear regression models, they found significant positive correlations between subjective flow scores and EEG power in the delta, theta, and gamma bands within specific time windows. A full-cap EEG setup was employed in [21] to assess EEG dynamics and the neural generators of flow. Event-related spectral perturbation (ERSP) was used to quantify time-locked changes in frequency bands during specific actions or performance bursts. The results indicated that the alpha power in the frontal and central regions varies with task demand; moderate increases are associated with the absorption and relaxed alertness characteristic of flow. ERSP and time-resolved analyses further revealed how these oscillatory patterns evolve at the onset and during the maintenance of flow episodes. A study investigated the neural correlates of flow using portable EEG devices and found significant correlations between EEG power in the delta, gamma, and theta bands and subjective flow scores, with predictive modeling conducted through regression and machine learning techniques [22]. According to [23] the increase in theta and delta power in EEG signals may serve as potential indicators of flow state. Despite encouraging results on flow state assessment with EEG, most studies employ dense, multi-channel configuration which increases the complexity of EEG device use for user. These configurations also require coverage of hair bearing scalp, which heightens susceptibility to noise and interference [11]. Overall, these limitations have impeded practical deployment in real-world settings.

The emergence of wearable, consumer-grade EEG headsets has created new avenues for identifying flow in everyday settings [21]. These platforms offer a practical means of tracking flow while mitigating the constraints associated with high-density, multi-channel EEG systems. A Muse EEG headband was used in [24] to investigate flow-state dynamics from prefrontal EEG recordings. Spectral power features were extracted during sports-related tasks (e.g., golf and tennis). The findings indicated modest predictive capability and highlighted the importance of frontal brain regions in flow. In another study researchers examined EEG correlates of flow state across various tasks (Tetris, mindfulness, art) using a single-channel prefrontal EEG system. The proposed method demonstrated that spectral power changes, especially in delta, theta, and gamma bands, correlate with subjective flow reports [20]. Despite their potential, wearable-based studies often cannot adequately filter real-life artifacts (blinks, muscle activity, environmental noise), yielding low-SNR data. Reported models frequently show modest predictive power, small samples that limit generalizability, and reliance on self-reported flow that can bias results. Consequently, consumer-oriented EEG remains underused for robust flow-state detection.

Thus, in this study we tested whether non-reciprocal co-activation of the sympathetic and parasympathetic systems can serve as a physiological marker of the flow state by analyzing electroencephalogram (EEG) recordings during Tetris gameplay and how they covaried with post-trial subjective flow ratings. Task difficulty (operationalized as required performance speed) was manipulated across three conditions (Easy, Optimal, Difficult) to elicit flow, given that flow typically occurs at moderate task demands, whereas too little or too much demand reduces its likelihood. To enhance signal quality, noise and artifacts were removed from the raw EEG using wavelet-based denoising, after which entropy-based features were extracted from EEG sub-bands. Finally, Random Forest (RF) classifier was trained to distinguish flow from relaxation. The proposed methodology was evaluated on two EEG datasets collected with consumer-oriented devices, supporting its suitability for real-world applications. The proposed framework, illustrated in Fig. 1, enables end-to-end flow state detection using consumer-grade EEG headbands through a four-stage pipeline.

## 2. Materials and methods

### 2.1. Study protocol and data acquisition

The flow state is a mental condition marked by intense concentration and complete immersion in an activity, occurring when a person's abilities are well-suited to the difficulty of the task [25]. Studies of flow during Tetris gameplay have identified several physiological and neural indicators. Because non-reciprocal co-activation of the sympathetic and parasympathetic systems may index the flow state, external factors that influence attention such as engagement level and visual stimuli (e.g., color and lighting) should be carefully controlled in experimental designs [26].

In this study, together with our previous work, [27], a modified version of a Tetris game was used to induce the psychological state of flow, defined by a high level of focus and full engagement in the task. First, participants were asked to play the game at three predetermined speeds. They then selected the speed they felt most effectively supported their experience of flow, as the pace of gameplay plays a crucial role in maintaining the delicate balance between boredom and overwhelming difficulty. Each participant subsequently completed three gameplay sessions, each lasting five minutes, with a one-minute rest with eyes closed between sessions to minimize fatigue and reduce potential carryover effects. Flow labels were assigned to EEG segments recorded during gameplay at the participant-selected optimal difficulty level, which was chosen after an initial calibration phase. This selection was based on participants' subjective assessment of the gameplay speed most conducive to experiencing flow. Relaxation labels corresponded to the 1-minute eyes-closed resting periods recorded immediately after each gameplay round. Thus, each participant contributed five gameplay segments (flow condition) and one relaxation segment per round. In this paper, we utilized two different consumer-oriented EEG devices, Muse-s EEG headband [28] and Emotiv Insight headset [29], to record EEG signals and validate the performance of the proposed method across various commercial, wearable equipment. Although the experimental setup was largely the same, each device had different configurations and sampling rates. Details of each dataset are provided below.

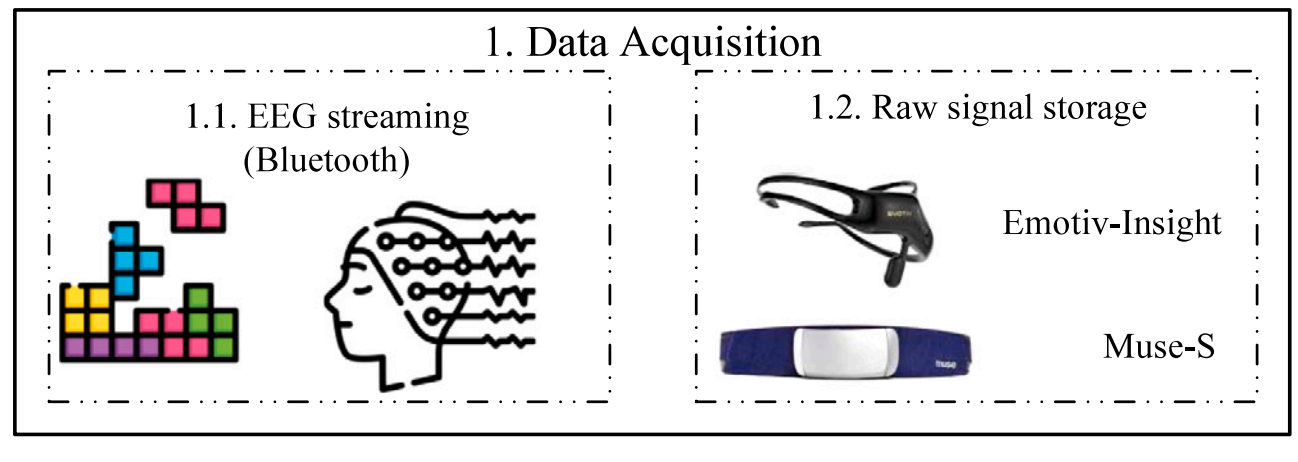

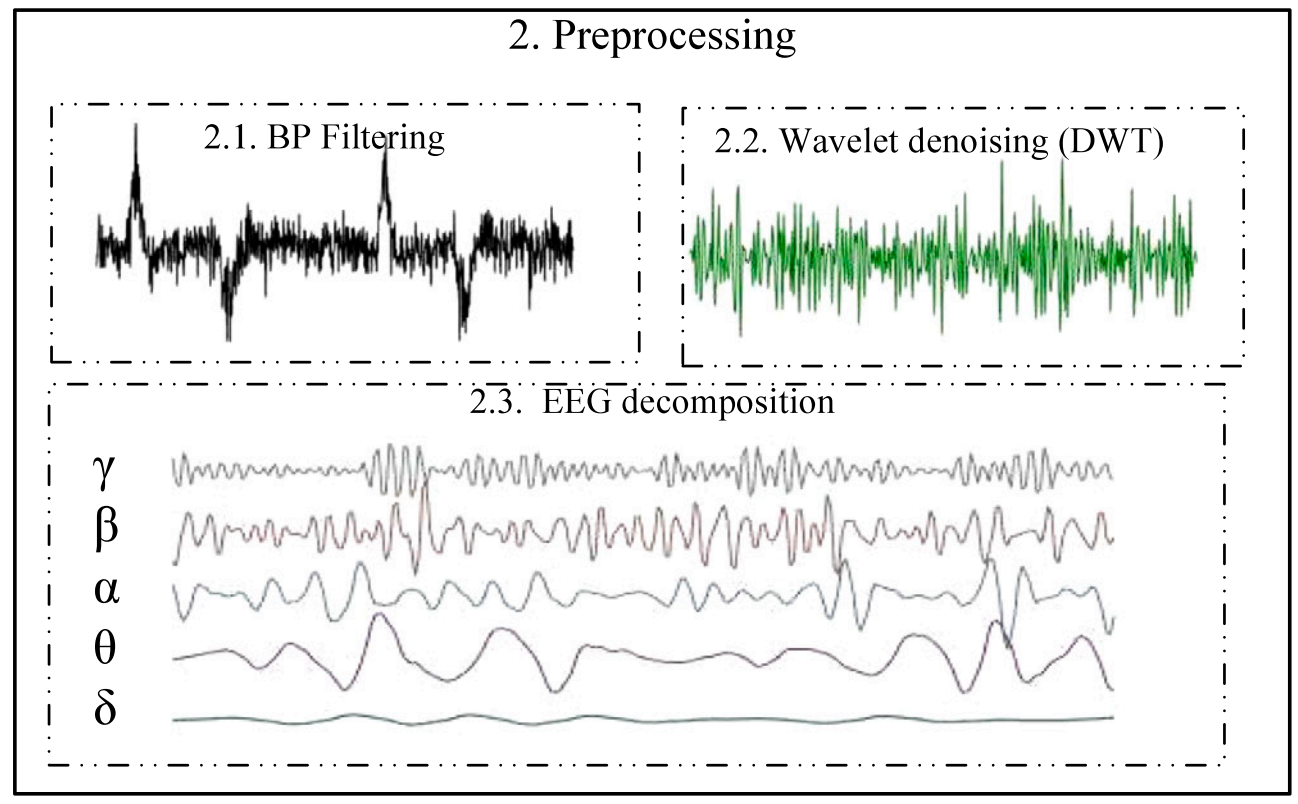

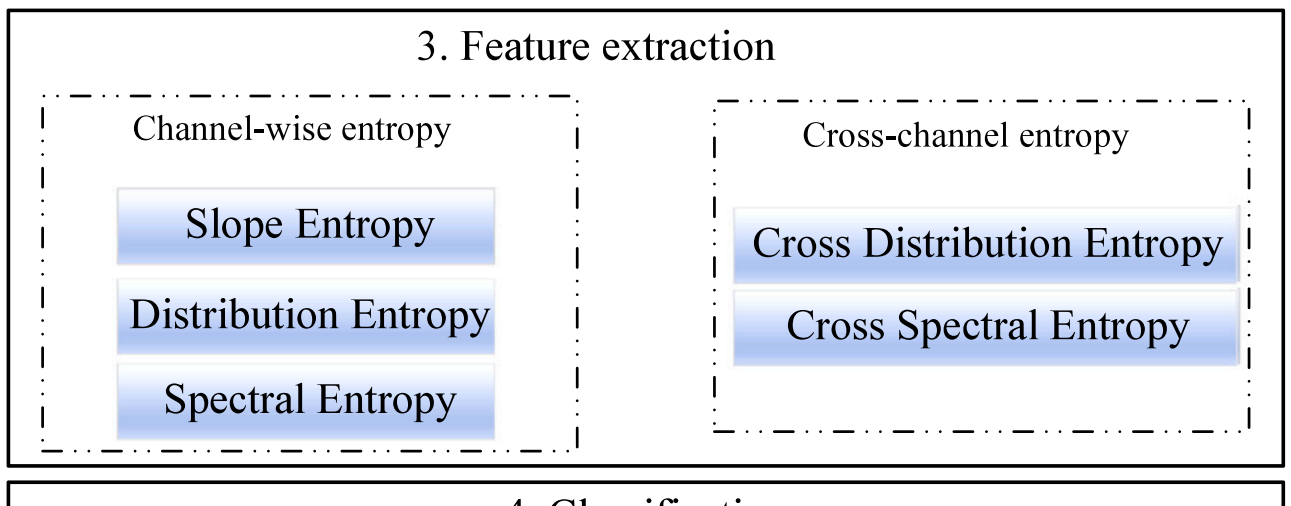

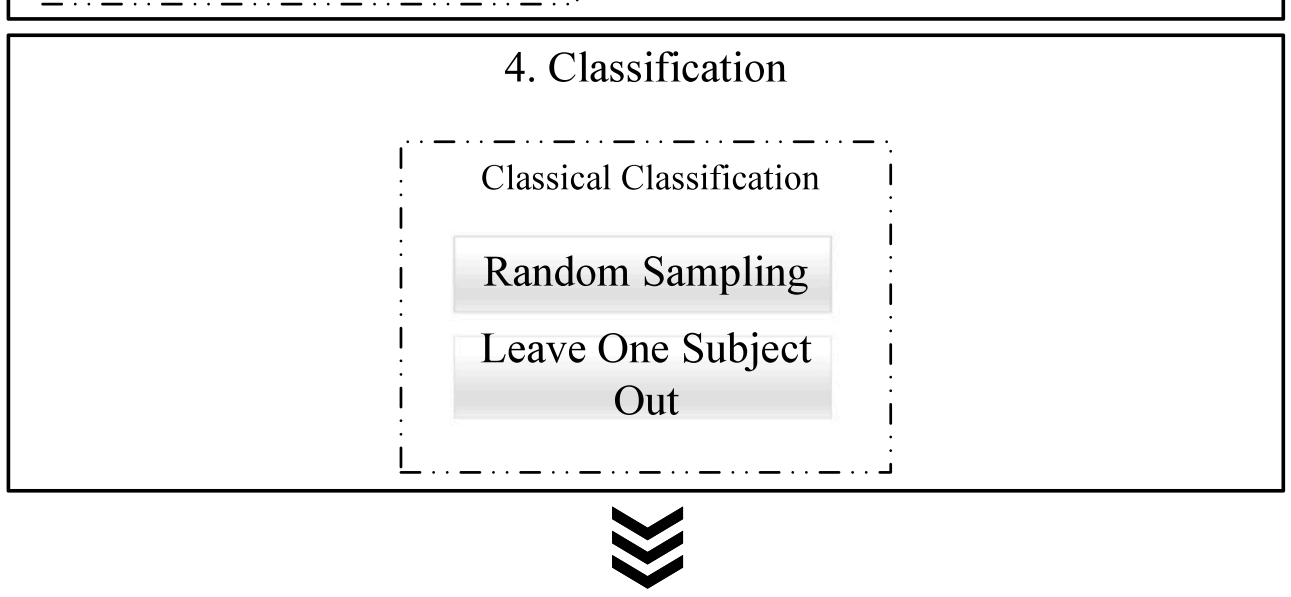

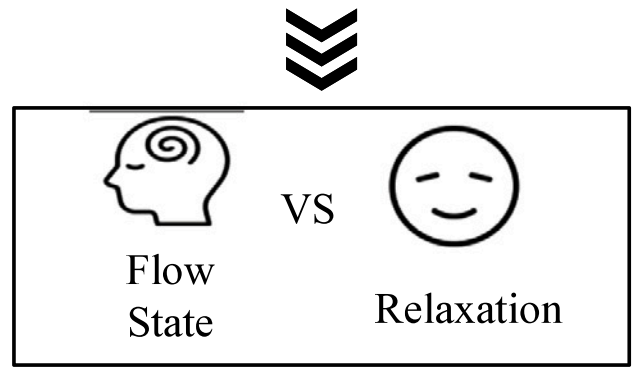

**Fig. 1.** The proposed framework for flow state detection using EEG headsets. The pipeline comprises four primary stages: (1) Data Acquisition, where EEG signals are streamed from the Muse-S or Emotiv Insight headset and stored locally. (2) Preprocessing, including a Butterworth band-pass filtering and the DWT denoising to remove artifacts. (3) EEG Decomposition and Feature Extraction, where signals are decomposed into frequency sub-bands and entropy-based features are computed, including channel-wise descriptors and cross-channel measures. (4) Classification, where extracted features are fed into multiple machine learning classifiers and evaluated under RS-CV and LOSO-CV schemes to distinguish flow from relaxation states.

### 2.2. *Data acquisition*

#### 2.2.1. *Muse data, data set M*

For the initial dataset, EEG was captured using a consumer-grade EEG device, the MUSE-S®, which integrates dry fabric-based electrodes located at AF7, AF8, TP9, and TP10. The Muse-S is designed for practical applications in everyday settings, as its dry sensors remove the necessity for conductive gels. In this work, signals from AF7 (Ch1) and AF8 (Ch2) were analyzed, as these sites are positioned on the prefrontal cortex and generally yield reduced impedance at the skin–electrode interface [28]. The reference electrode was fixed at Fpz. Data logging was managed with external software provided by Petal Technology LLC [30] and executed on a personal computer. EEGs were streamed wirelessly from the Muse-S headband to the PC through a Bluetooth link, sampled at 256 Hz. Consistent electrode contact with participants' skin was maintained throughout the sessions, ensuring dependable data quality. The participant pool included 29 participants, aged between 18 and 68 years, offering a broad demographic for investigating gameplay-related flow states.

#### 2.2.2. *Emotive insight data, dataset E*

Following the successful acquisition of the Muse dataset, a comparable study was carried out using another commercially available EEG device, the Emotiv Insight [29]. This headset incorporates five semi-dry hydrophilic polymer electrodes, making it suitable for real-life settings as well as brain–computer interaction (BCI) research. Unlike conventional wet electrodes, the hydrophilic sensors function without conductive gel. Instead, the manufacturer advises applying a saline-based solution to sustain electrical conductivity. The device is equipped with electrodes distributed at AF3, AF4, T7, T8, Pz, together with two reference points (CMS/DRL) placed at the left mastoid region. Signal capture was performed using proprietary software supplied by Emotiv and operated on a desktop computer. EEGs were transmitted wirelessly from the Insight to the PC over a Bluetooth channel, with a sampling rate of 128 Hz. Secure electrode–skin contact was preserved throughout the sessions, ensuring robust and consistent recordings. Although the prefrontal electrodes lie over lateral prefrontal regions involved in executive control and attentional focusing, so that entropy changes there reflect modulation of key nodes within the broader hypofrontality network, these anterior frontal sites do not cover medial and dorsal PFC and are also closer to ocular and facial muscles. This dataset included 16 participants. Ethical clearance for both datasets was obtained from the Human Sciences Ethics Committee of the Universities in Satakunta, Finland (approval ID: 17.12.2024). Fig. 1 depicts the proposed pipeline for detecting flow state in a game-based environment. The algorithm comprises four stages: EEG data acquisition, preprocessing, feature extraction, and classification. The following subsections describe each stage in detail.

### 2.3. Preprocessing and artifact removal

An initial zero-phase Butterworth band-pass filter (0.5–40 Hz) was applied to attenuate low-frequency drift and high-frequency noise while preserving physiologically relevant EEG components. Artifacts were then removed using discrete wavelet transform (DWT) with hard thresholding, targeting ocular blinks, muscle activity, and environmental noise. Consistent with our previous work [31], we selected the db4 mother wavelet due to its morphology's close match to eye-blink dynamics. The DWT decomposition was performed hierarchically across six levels for the dataset M (256 Hz) and five levels for the dataset E (128 Hz), with adaptive thresholds calculated from median absolute deviation applied to detail coefficients. The artifact-corrected signal was reconstructed via inverse DWT, effectively removing contamination while preserving neural signatures relevant to flow-state detection.

### 2.4. EEG subband decomposition

Each channel of EEG signal was initially decomposed into an approximation component $a_1[n]$ and a detail component $d_1[n]$ using Discrete Wavelet Transform (DWT). The approximation component $a_1[n]$ was subsequently decomposed into the second-level approximation $a_2[n]$ and detail $d_2[n]$ components. The decomposition was iterated up to the maximum DWT level $L$. Consequently, the original signal $x[n]$ can be represented as the sum of the final approximation and all detail components:

$$x[n] = \sum_{l=1}^{L} d_l[n] + a_L[n], \tag{1}$$

The frequency band of each approximation and detail component can be obtained by:

$$a_l = \left[0, \frac{Fs}{2^{l+1}}\right], d_l = \left[\frac{Fs}{2^{l+1}}, \frac{Fs}{2^l}\right], \tag{2}$$

Where $Fs$ denotes the sampling rate. To eliminate eye-blink artifacts, the final approximation component $a_6$ for Muse Data and $a_5$ for Insight Data was denoised using adaptive thresholds, and the extracted noise component was subtracted from the original signal. In the following analysis, we adhere to the standard EEG frequency bands: $\delta, \theta, \alpha, \beta$ and $\gamma$. Given the sampling frequencies of 256 Hz for Dataset M and 128 Hz for Dataset E, the resulting frequency intervals approximately corresponded to the classical EEG bands as follows: $\delta$ (0–4 Hz), $\theta$ (4–8 Hz), $\alpha$ (8–16 Hz), $\beta$ (16–32 Hz), and $\gamma$ (32–64 Hz).

### 2.5. Feature extraction

Entropy features quantify the irregularity and complexity of EEG signals at the levels where cognition varies, across frequency bands, time scales, and channels. Different types of entropy have been widely used in research on detecting diverse mental states, and they have shown to be powerful tools for identifying the flow state and relaxation.

#### 2.5.1. Channel-wise entropy

Following DWT-based decomposition of the EEG into sub-bands, we derived three complexity descriptors: Slope Entropy (SlopEn), Spectral Entropy (SpectEn), and Distribution Entropy (DisEn). SlopEn characterizes the temporal steepness of amplitude fluctuations how quickly the signal changes and where lower values generally reflect more ordered, synchronized neural activity; accordingly, it has been used to monitor anesthetic state and to discriminate mental conditions such as flow. The operational parameters for SlopEn were set in accordance with [32]. SpectEn quantifies the dispersion of power across frequencies via the entropy of the normalized spectrum and has been widely reported as a robust biomarker in anesthesia EEG; it has also been applied to pediatric EEG in Autism Spectrum Disorder (ASD), particularly over frontal and temporal regions, revealing patterns distinct from neurotypical controls. Nevertheless, its potential for flow-state detection remains largely unexplored [33]. Distribution Entropy (DisEn) captures signal unpredictability by estimating sub-band probability distributions [34]; in this study we used an embedding dimension of m=2, a time delay of $\tau$=1, and a logarithm base of 2.

#### 2.5.2. Cross-channel entropy

Cross-channel entropy quantifies nonlinear, time varying dependencies among EEG channels. By characterizing joint complexity and interdependence, it provides a window into network-level processes underlying cognitive states such as workload, attention, emotion, and flow. Compared with univariate entropy measures, cross-channel variants more faithfully index functional connectivity and large-scale brain dynamics [35]. Cross distribution entropy, often referred to simply as cross-entropy, is a measure from information theory that quantifies the dissimilarity or difference between two probability distributions over the same set of events. In mathematical terms, given two probability distributions $p$ (true distribution) and $q$ (estimated distribution), the cross-entropy $H(p, q)$ measures the number of bits needed to represent or encode samples from distribution $p$ when using a coding scheme optimized for $q$.

$$H(p, q) = -\sum_{x\in X} p(x) \log q(x) \tag{3}$$

where $p(x)$ is the true probability of event $x$ and $q(x)$ is the predicted probability [36].

#### 2.5.3. Feature dimensionality across datasets

The two EEG headsets used in this study differed in channel configuration. Dataset M (Muse-S) contained two prefrontal channels (AF7 and AF8), whereas Dataset E (Emotiv Insight) contained five channels (AF3, AF4, T7, T8, and Pz). For each channel, three entropy measures, namely Slope Entropy (SlopEn), Distribution Entropy (DisEn), and Spectral Entropy (SpecEn), were extracted from five EEG sub-bands ($\delta$, $\theta$, $\alpha$, $\beta$, and $\gamma$). Consequently, Dataset M produced 30 channel-wise features ($2 \times 5 \times 3$), while Dataset E produced 75 channel-wise features ($5 \times 5 \times 3$). For cross-channel analysis, entropy features were computed for all unique channel pairs within each dataset. As the number of channel pairs depends on the available electrodes, the dimensionality of the cross-channel feature set also differed between datasets. To avoid introducing bias related to feature-space dimensionality, classification models were trained and evaluated independently for each dataset. Therefore, no feature-space alignment, feature selection, dimensionality reduction, or feature matching between the Muse-S and Emotiv Insight datasets was performed. Comparisons between devices were conducted at the level of classification performance rather than through joint model training.

### 2.6. Classification methods

A variety of machine learning methods have been explored for distinguishing flow and non-flow states using physiological and behavioral data from multiple sources. These approaches range from traditional classifiers to advanced models, consistently demonstrating promising accuracy in detecting the flow experience across diverse experimental paradigms. To improve reproducibility, the principal hyperparameter settings of all classifiers were explicitly defined. To ensure a fair comparison among classifiers and to isolate the contribution of the extracted entropy features, identical classifier configurations were maintained across all datasets and validation schemes. No dataset-specific hyperparameter optimization was performed. Consequently, the reported results should be interpreted as an evaluation of the discriminative capability and generalizability of the proposed entropy features rather than the maximum achievable performance of each classifier under extensive tuning.

#### 2.6.1. Random Forest

Random Forest (RF) models have been widely leveraged to infer mental conditions from electroencephalographic (EEG) recordings. Previous work has documented strong performance in estimating stress [14], affective states [37], and cognitive fatigue [38]. RF is an ensemble method that aggregates many decision trees to enhance generalization and temper variance: Each tree is trained on a bootstrap resample of the data (bagging), often with randomized feature selection in splits, and the ensemble's decision is obtained by majority vote. In our implementation, we employed 100 trees with unlimited depth `max_depth=None` . This setup delivers competitive accuracy, resilience to measurement noise, and a reduced tendency to overfit compared to single-tree classifiers.

#### 2.6.2. k-Nearest Neighbors

The k-Nearest Neighbors (kNN) algorithm represents a fundamental and widely-used approach to classification tasks. The algorithm operates by identifying the k closest training samples to a given input and assigning a class label based on the predominant class among these neighboring points. A critical parameter in kNN is the selection of k, which directly influences the complexity of the decision boundary. Lower k values produce more intricate decision boundaries that may be susceptible to overfitting by capturing underlying noise in the data, whereas higher k values yield smoother, more generalized boundaries that better represent broader patterns in the dataset. The k-nearest neighbors classifier was implemented using $k = 5$ and Euclidean distance.

#### 2.6.3. Support Vector Machines

Support Vector Machines (SVM) constitute a robust classification methodology that identifies an optimal separating hyperplane within high-dimensional feature spaces to effectively distinguish between distinct classes. The fundamental objective of SVM is to maximize the margin between class boundaries, thereby improving the model's capacity for generalization to unseen data. A notable strength of SVMs lies in their versatility to address both linearly separable and non-linearly separable datasets through the application of kernel functions, which map the original input space into higher-dimensional representations where linear class separation becomes feasible. The most recommended kernels to be used in SVM methods in research are RBF and linear kernels [39]. For the support vector machine classifiers, the RBF-SVM employed a radial basis function kernel with automatic kernel scaling and default box-constraint settings, while the linear SVM used a linear kernel.

#### 2.6.4. GentleBoost

Boosting is an ensemble learning technique that sequentially combines weak learners into a strong learner by iteratively adjusting sample weights to focus on misclassified instances. Prominent variants include GentleBoost, Gradient Boosting, and XGBoost. GentleBoost also known as Gentle AdaBoost, combines features of AdaBoostM1 and LogitBoost while employing conservative, adaptive Newton steps rather than aggressive reweighting. This approach renders GentleBoost less sensitive to noise and outliers compared to standard AdaBoost, making it particularly suitable for real-world scenarios with noisy or mislabeled data. The smoother optimization strategy of GentleBoost provides enhanced stability and generalization compared to more aggressive boosting variants. In our implementation GentleBoost was trained using 100 weak learners [40].

#### 2.6.5. Fitted discriminant classifiers

Fitted discriminant classifiers derive linear decision boundaries by modeling class-conditional probability distributions, typically assuming multivariate Gaussian distributions. They maximize the ratio of between-class to within-class variance, effectively identifying linear feature combinations that separate distinct classes. While assuming multivariate normality and equal covariance structures across classes, extensions such as Quadratic Discriminant Analysis (QDA) and regularized discriminant analysis accommodate class-specific covariances and high-dimensional settings. In biomedical applications, fitted discriminant classifiers have achieved competitive accuracy in EEG signal classification and brain–computer-interface systems. The fitted discriminant classifier employed a linear discriminant model [41].

#### 2.6.6. Naive Bayes classifier

The Naive Bayes classifier is a probabilistic algorithm that applies conditional independence assumptions to compute posterior class probabilities via maximum a posteriori (MAP) decision rule. Despite oversimplified independence assumptions, it demonstrates remarkable effectiveness with minimal training data and computational requirements, making it advantageous for high-dimensional datasets and real-time applications. In biomedical signal processing, Naive Bayes has successfully classified EEG signals for neurological disorder detection and brain–computer interfaces, often competing favorably with more computationally complex methods. The Naive Bayes classifier used Gaussian class-conditional distributions [42].

## 3. Evaluation

All features derived from the artifact-free dataset were subjected to min–max normalization, scaling each feature to the range. Following normalization, the data was partitioned by randomly selecting 80% of the feature vectors for model training and validation, while the remaining 20% served as the independent test set. To ensure robust assessment of classification performance, the train–test split and evaluation procedure were iterated over 100 randomized runs. The following criteria were calculated to evaluate the performance of the classifiers:

$$\text{Accuracy} = \frac{T_P + T_N}{T_P + T_N + F_N + F_P} \times 100, \tag{4}$$

$$\text{Precision} = \frac{T_P}{T_P + F_P} \times 100, \tag{5}$$

$$\text{Recall} = \frac{T_P}{T_P + F_N} \times 100, \tag{6}$$

$$\text{F1-score} = \frac{T_P}{T_P + (F_P + F_N) \times \frac{1}{2}} \times 100, \tag{7}$$

where $T_P$, $T_N$, $F_P$ and $F_N$ denote the number of true positives, true negatives, false positives and false negatives, respectively.

## 4. Results

Each participant completed three gameplay rounds, each lasting five minutes. We trained machine learning methods to distinguish game segments from relaxation periods within each session and to examine temporal patterns in feature dynamics across the recording. Each round was partitioned into five 1-minute gameplay segments plus a 1-minute relaxation block; thus, per session (three rounds), the models received three sets of five 1-minute game segments and three 1-minute relaxation segments. To assess generalizability and robustness, we evaluated performance under two cross-validation schemes: Random Sampling (RS) and Leave-One-Subject-Out (LOSO). For RS on Databases E and M, Tables 1 and 2 report the Random Forest results. To aid interpretation, we ran experiments 100 times using the full feature vectors, 30 features for Database M (2 channels × 5 sub-bands ×3 features) and 75 features for Database E (5 channels × 5 sub-bands ×3

**Table 1**
Output of the Random Forest classifier for random sampling cross-validation for Dataset E. Mean ± SD over 100 runs.

| Features | Metric | 1st | 2nd | 3rd | 4th | 5th |
|---|---|---|---|---|---|---|
| All entropies | Accuracy | 97.23 ± 1.02 | 97.06 ± 1.02 | 96.39 ± 1.02 | 97.94 ± 1.02 | 97.32 ± 1.02 |
| | Precision | 97.57 ± 0.80 | 98.04 ± 0.69 | 96.65 ± 0.71 | 97.76 ± 0.71 | 97.41 ± 0.44 |
| | Recall | 96.24 ± 1.08 | 96.66 ± 1.88 | 96.38 ± 1.48 | 97.06 ± 1.84 | 97.18 ± 1.10 |
| | F1-Score | 97.18 ± 2.37 | 97.33 ± 2.17 | 96.47 ± 1.88 | 97.39 ± 1.29 | 97.28 ± 1.48 |
| SlopeEn | Accuracy | 90.26 ± 1.17 | 89.52 ± 1.37 | 91.02 ± 1.07 | 91.28 ± 1.62 | 90.39 ± 1.02 |
| | Precision | 89.79 ± 1.71 | 89.61 ± 1.30 | 90.21 ± 1.49 | 90.41 ± 1.73 | 90.79 ± 1.13 |
| | Recall | 90.86 ± 0.83 | 89.33 ± 0.96 | 92.28 ± 0.75 | 92.29 ± 0.89 | 89.69 ± 1.41 |
| | F1-Score | 90.26 ± 1.67 | 88.39 ± 1.71 | 91.15 ± 1.40 | 91.32 ± 1.48 | 90.33 ± 1.48 |
| DisEn | Accuracy | 91.36 ± 1.02 | 92.40 ± 1.02 | 92.84 ± 1.02 | 94.19 ± 1.02 | 92.18 ± 1.02 |
| | Precision | 92.76 ± 1.96 | 92.70 ± 2.35 | 93.18 ± 1.96 | 93.24 ± 1.82 | 92.50 ± 1.70 |
| | Recall | 90.59 ± 0.79 | 92.08 ± 0.92 | 92.39 ± 0.72 | 95.33 ± 0.87 | 91.59 ± 1.03 |
| | F1-Score | 91.60 ± 1.55 | 92.31 ± 1.57 | 92.71 ± 2.10 | 94.24 ± 2.09 | 92.17 ± 1.76 |
| SpecEn | Accuracy | 94.25 ± 1.02 | 95.13 ± 1.02 | 95.42 ± 1.02 | 95.09 ± 1.02 | 95.34 ± 1.02 |
| | Precision | 94.42 ± 0.34 | 96.66 ± 0.30 | 96.80 ± 0.56 | 96.41 ± 0.85 | 96.67 ± 0.82 |
| | Recall | 95.66 ± 2.10 | 95.10 ± 2.03 | 95.38 ± 2.53 | 94.95 ± 1.34 | 94.76 ± 2.11 |
| | F1-Score | 95.02 ± 1.53 | 95.84 ± 2.01 | 96.36 ± 1.86 | 95.65 ± 1.25 | 95.67 ± 1.31 |

RS: random sampling; RF: Random Forest. Bold indicates the best value within each entropy type.

**Table 2**
Output of the Random Forest classifier for random sampling cross-validation for Dataset M. The total accuracy is given as the mean ± standard deviation over the 100 random sampling sets.

| Features | Metric | 1st | 2nd | 3rd | 4th | 5th |
|---|---|---|---|---|---|---|
| All entropies | Accuracy | 93.12 ± 1.51 | 93.13 ± 1.47 | 92.98 ± 1.51 | **93.26 ± 1.58** | 91.62 ± 1.58 |
| | Precision | 89.96 ± 1.02 | 89.92 ± 1.52 | 89.85 ± 1.30 | 89.99 ± 1.79 | 88.81 ± 1.32 |
| | Recall | 99.42 ± 2.01 | 99.47 ± 2.25 | 99.54 ± 1.92 | 99.41 ± 2.19 | 99.39 ± 2.02 |
| | F1-Score | 94.43 ± 2.00 | 94.44 ± 2.02 | 94.53 ± 2.19 | 94.45 ± 1.92 | 93.78 ± 2.08 |
| SlopeEn | Accuracy | 84.06 ± 1.23 | 83.53 ± 1.31 | **84.37 ± 1.45** | 82.80 ± 1.37 | 81.60 ± 1.40 |
| | Precision | 83.22 ± 1.82 | 81.48 ± 1.64 | 83.10 ± 1.12 | 82.61 ± 1.53 | 80.78 ± 1.82 |
| | Recall | 92.49 ± 0.61 | 92.35 ± 0.85 | 92.34 ± 0.92 | 91.90 ± 0.97 | 90.66 ± 1.06 |
| | F1-Score | 87.54 ± 0.94 | 86.52 ± 0.94 | 87.43 ± 0.92 | 86.95 ± 1.07 | 85.38 ± 1.10 |
| DisEn | Accuracy | 92.72 ± 1.67 | **93.14 ± 1.55** | 93.07 ± 1.60 | 92.76 ± 1.52 | 92.39 ± 1.37 |
| | Precision | 89.79 ± 1.24 | 89.86 ± 1.13 | 89.94 ± 1.54 | 90.20 ± 1.14 | 88.34 ± 1.39 |
| | Recall | 99.04 ± 1.57 | 99.46 ± 1.14 | 99.31 ± 1.39 | 99.53 ± 1.13 | 99.46 ± 1.26 |
| | F1-Score | 94.16 ± 1.33 | 94.10 ± 1.25 | 94.61 ± 1.23 | 94.62 ± 1.09 | 93.55 ± 1.18 |
| SpecEn | Accuracy | 91.08 ± 1.88 | 91.28 ± 2.21 | 91.56 ± 1.76 | **91.73 ± 1.65** | 91.00 ± 1.82 |
| | Precision | 88.99 ± 1.86 | 89.02 ± 1.55 | 89.39 ± 1.66 | 88.80 ± 1.84 | 88.00 ± 1.84 |
| | Recall | 98.15 ± 1.62 | 98.26 ± 1.46 | 98.34 ± 1.79 | 98.25 ± 1.79 | 98.15 ± 1.75 |
| | F1-Score | 93.33 ± 1.96 | 93.21 ± 1.52 | 93.64 ± 2.01 | 93.64 ± 1.79 | 92.84 ± 1.75 |

RS: random sampling; RF: Random Forest. Bold indicates the best value within each row.

features). Random-sampling cross-validation consistently outperformed LOSO across both datasets, yielding higher classification accuracy and more reliable estimates (see Tables 3 and 4). For Dataset E, across the five 1-min segments, the all entropies model achieved the highest accuracy (~96.4%–97.5% ± 1.0–1.1), with precision/recall/F1-Score in the 96%–98% range and a peak around Segment 4. Among single-feature families, spectral entropy (SpecEn) performed best (~94.7%–96.4%), distribution entropy (DisEn) was mid-tier (~91%–94%), and slope entropy (SlopEn) lagged (~89%–91%). Variation across segments within a feature set was small, and metric dispersion was tight, indicating stable training. For Dataset M (Table 4), All entropies again led with ~91.6%–93.3% accuracy (±1.5–1.6), highest near Segment 4. DisEn was a close second (~92%–93%), SpecEn slightly lower (~91%–92%), and SlopEn the weakest (~81%–84%). Sensitivity for the all-entropy models was uniformly high (≈99%), whereas specificity was lower (≈83%), indicating most errors came from negatives (rest epochs were classified as gaming sessions). Standard deviations were modest, again suggesting stable estimates.

Across both datasets, combining the three entropy groups, all entropies yields the best or tied-best results, and performance varies only weakly across the 1-min segments, with Segment 4 most often the top performer. Following the encouraging classifier results with channel-wise entropy features, we also computed cross-channel entropy features such as cross-spectral entropy and cross-distribution entropy as described in Section 2.

**Table 3**
Output of the Random Forest classifier for all channel-wise entropy features and LOSO-CV on Dataset E. Reported values are mean ± SD across subjects.

| EEG segments | Accuracy | Precision | Recall | F1-Score |
|---|---|---|---|---|
| 1st & Relaxation | **75.65 ± 1.22** | 76.09 | 77.77 | 75.86 |
| 2nd & Relaxation | 74.47 ± 1.37 | 78.37 | 77.43 | 75.26 |
| 3rd & Relaxation | 73.26 ± 1.14 | 76.54 | 75.34 | 74.06 |
| 4th & Relaxation | 73.09 ± 1.05 | 75.65 | 75.69 | 73.62 |
| 5th & Relaxation | 71.52 ± 1.26 | 73.45 | 76.73 | 73.44 |

**Table 4**
Output of the Random Forest classifier for all channel-wise entropy features and LOSO-CV on Dataset M. Reported values are mean ± SD across subjects.

| EEG segments | Accuracy | Precision | Recall | F1-Score |
|---|---|---|---|---|
| 1st & Relaxation | **82.75 ± 1.02** | 81.89 | 89.84 | 83.91 |
| 2nd & Relaxation | 78.16 ± 1.17 | 80.28 | 85.05 | 80.37 |
| 3rd & Relaxation | 82.18 ± 1.25 | 81.93 | 85.82 | 82.08 |
| 4th & Relaxation | 79.42 ± 0.95 | 80.90 | 86.39 | 81.66 |
| 5th & Relaxation | 80.42 ± 1.31 | 81.25 | 87.41 | 82.13 |

Beyond channel-wise features, we included cross-channel entropy measures (cross-spectral and cross-distribution entropy) to capture functional coupling and network-level dynamics that are often more stable across individuals than single-channel irregularity. Cross-channel entropy was extracted between each subband of each pair of channels as it was explained in Section 2.5.2 in both datasets. For Dataset E, accuracy ranged from 75.2%–78.4% across the five 1-min segments (mean 76.7%), peaking at Segment 2 (78.40%), with Segment 4 (77.91%) close behind. Precision was 73.6%–75.9%, recall 80.7%–84.1%, and F1-Score 76.6%–79.4%, with modest segment-to-segment variability. For Dataset M, accuracy was uniformly higher, 79.3%–81.4% (mean 80.7%), peaking at Segment 3 (81.43%); precision was 78.1%–79.5%, recall 82.1%–85.3%, and F1-Score79.9%–82.0%, again with small variability. The dataset M exceeded dataset E in accuracy at every segment by 3.15, 2.93, 5.44, 2.87, and 5.67 percentage points, respectively. Performance varied only slightly across segments, with the middle segments (2–4) generally strongest (see Fig. 2). Overall accuracy in Dataset M is higher in every segment (≈79.3%–81.4%) than in Dataset E (≈75.2%–78.4%), with the largest gaps in Segments 3 and 5 (≈5–6 percentage points). Precision in Dataset M remains around 78.1%–79.5%, whereas Dataset E is lower at 73.2%–75.9%, again showing the biggest advantages in the mid-to-late session. Both datasets achieve strong recall, but Dataset M (≈82.1%–85.3%) edges Dataset E (≈80.7%–84.1%) across all segments, with Segment 4 the closest. The F1-Score advantage also persists (Dataset M ≈ 79.9%–82.0% vs. Dataset E ≈ 76.6%–79.4%), mirroring the precision/recall pattern (Fig. 6).

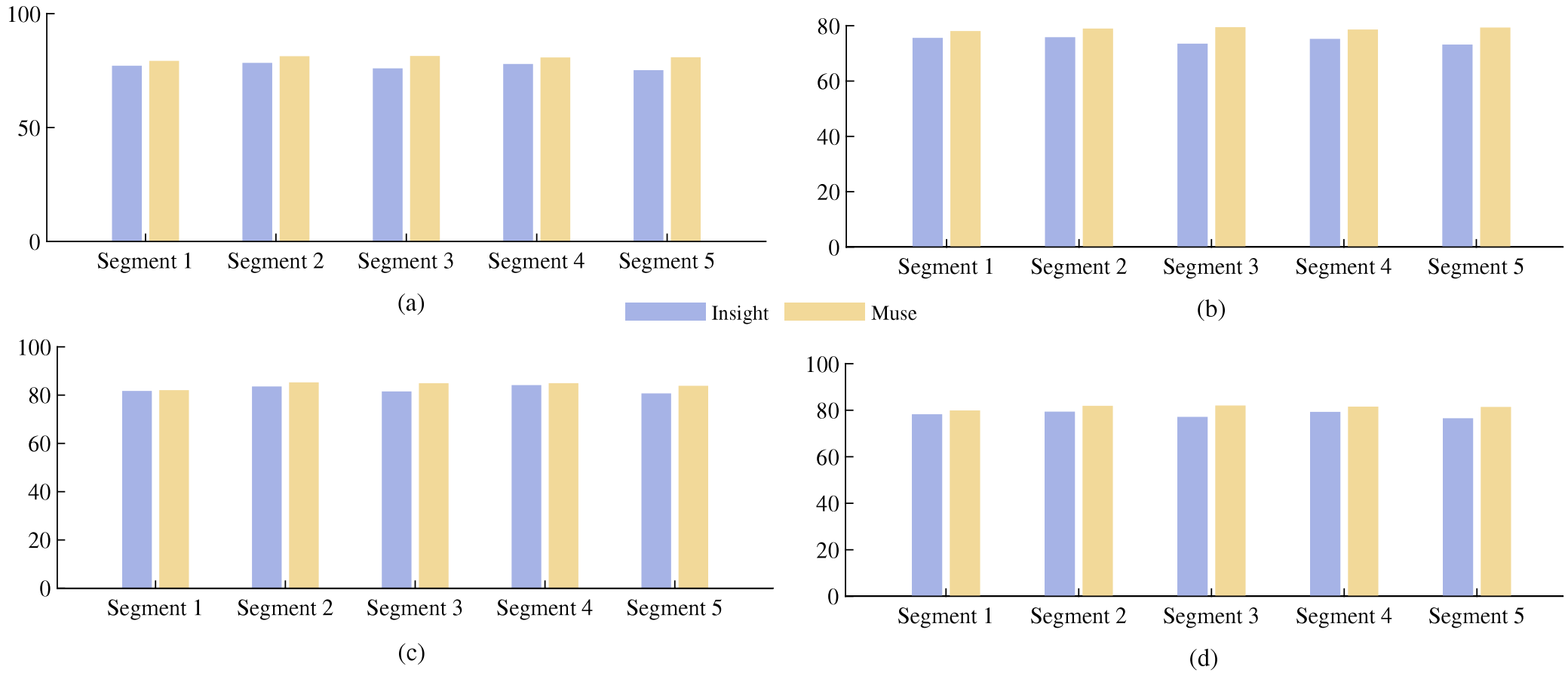

**Fig. 2.** Classification performance metrics for cross-entropy features ((a): accuracy, (b): precision, (c): recall, and (d): F1-score) comparing Dataset M (Muse) and Dataset E (Emotiv) across five consecutive 1-minute gameplay segments, showing consistent mid-session performance peaks and stable reliability of entropy-based features for cross-device flow detection.

Across all five segments, dataset M outperforms Dtase E on accuracy, precision, recall, and F1-Score by about 2–5 percentage points with the largest margins around Segments 2–3 (Fig. 6). Performance is fairly stable over time: mid-session segments peak (accuracy ≈81%–81.5% for dataset M vs. 76%–78% for dataset E), while Segment 5 is slightly lower for both; recall is consistently higher than precision, so F1-score sits between. While RS-CV reached 97% accuracy, the more rigorous LOSO evaluation achieved up to 82.75% accuracy, demonstrating meaningful cross-subject generalization. LOSO-CV is widely recommended for wearable EEG because it prevents identity leakage and directly probes robustness to inter-subject variability arising from head anatomy, electrode placement, impedance, and idiosyncratic cognitive strategies. Consistent with prior EEG literature, LOSO-CV typically yields lower accuracies than RS. RS can be optimistic when trials from the same participant appear in both train and test folds so we treat LOSO-CV as the more conservative estimate of out-of-subject generalization (See Table 3 and 4). Motivated by the strong performance under RS-CV, we also evaluated a stricter, subject-independent setting leave-one-subject-out cross-validation (LOSO-CV), on both datasets and for both feature groups (channel-wise and cross-channel entropies). Per-subject results are visualized and compared in Fig. 3. Across the LOSO-CV for Dataset E, subject-level accuracy generally clusters in the mid-to-high 70s with many values in the 80%–90% range and several above 90%. The best accuracies reach 97.22% (seen in Segments 2–4), while the lowest accuracies are about 55.56% (appearing in Segments 1, 3, and 5), indicating notable inter-subject spread. Precision is the most variable metric, ranging from about 53% to 100%, whereas recall typically stays higher, spanning about 44% to 100%; this imbalance keeps F1-Score between about 52% and 97% (with multiple F1-Score peaks around 97% in Segments 3–4). Taken together, the mid-session segments (3–4) yield the most top-end results (several subjects with 97% accuracy and F1-Score close to 97%), while Segments 1 and 5 show more low outliers. Overall, LOSO performance suggests good cross-subject generalization for many participants but also substantial inter-subject variability, driven primarily by swings in precision. The same strategy was applied to the Muse dataset, and the results are shown in Fig. 3.

Across subjects and segments, accuracies are generally in the mid 70s to low 90s, with several strong cases near 97% (e.g., Subject 2 in Segments 2 and 4), while the lower end appears around 55%–65% for a few subjects/segments (e.g., Subject 10 in Segment 4 at ~61% accuracy with precision ~57% and recall ~89%). Precision typically ranges ~75%–95%, and recall is often ~83%–100%, so the F1-score falls between the two and remains robust for many subjects (e.g., Subject 13 frequently shows precision = 100% with recall = 72%–100%, yielding F1-Score in the ~84%–94% band). Performance is fairly stable across the five 1-minute segments, with the mid-session windows (Segments 2–4) most often yielding the highest accuracies and F1s, whereas a few outliers create localized dips. Overall, the results indicate consistent, high recall and solid F1-Score for most subjects, moderate inter-subject variability, and the strongest results concentrated in Segments 2–4.

In the leave-one-subject-out evaluation, the Random Forest performance is summarized in Table 3 (Dataset E) and Table 4 (Dataset M). In both datasets, the highest accuracy occurs in the first 1-min game segment versus relaxation: 75.65% for Dataset E and 82.75% for Dataset M. Accuracy then declines modestly across subsequent segments. The same trend is visible in the other metrics: for Dataset E, precision, recall, and F1-score begin at 76.09%/77.76%/75.86% and drift slightly downward by the fifth segment; for Dataset M, the initial values are 81.89%/89.84%/83.91% and remain higher overall. In absolute terms, Dataset M outperforms Dataset E on all four criteria in every segment, yet both tables exhibit a consistent temporal profile with peak performance in Segment 1 followed by a gradual decline toward later segments.

We also applied RS-CV across EEG sub-bands in both datasets and the results are shown in Figs. 4 and 5. Across all five 1-min segments in dataset E, the ranking of EEG bands is highly stable: Gamma > Theta > Alpha > Beta > Delta. For example, in Segment 1 the accuracies are approximately Gamma 93.9%, Theta 89.9%, Alpha 88.3%, Beta 86.0%, and Delta 82.3%, with F1-scores mirroring this order. The same pattern holds in Segments 2–5, with gamma-band accuracy ~94% (F1-Score~92%–93%); theta ~89%–90%; alpha ~88%; beta ~85%–86%; and delta ~80%–82%. Within-band variability across segments is small (typically a few tenths of a percentage point), indicating stable estimates. The gamma–theta gap is roughly 3–5 percentage points (pp), and gamma exceeds delta by about 12–14 pp, underscoring the greater discriminative power of high-frequency activity for this task. Precision and recall follow the same ordering, yielding F1-Score values consistent with accuracy. (See Fig. 4.) As shown in Fig. 5, across all five 1-min segments in Dataset M, the band ranking is highly consistent. For example, in Segment 1 the accuracies are Delta 90.16%, Gamma 82.48%, Beta 79.77%, Alpha 68.24%, and Theta 70.37%; F1-scores mirror this order (Delta 90.85%, Gamma 83.36%, Beta 80.87%, Alpha 69.37%, Theta 70.59%). The same pattern holds in Segments 2–5: delta accuracy remains tightly clustered around ~90%–90.4% (F1-Score~90.9%–91.1%), gamma around ~80.7%–83.2% (F1-Score~81.7%–84.1%), beta ~77%–79% (F1-Score~78%–80%), and alpha/theta ~66%–71% (F1-Score~ 66%–69%). Within-band variability across segments is small especially for delta (range ≈0.7 pp) indicating stable estimates. Notably, delta exceeds gamma by ~6.5–9.6 pp and outperforms alpha/theta by ∼20+ pp, underscoring the stronger discriminative value of low-frequency (delta) activity for this dataset. Precision and recall follow the same ordering, yielding consistent F1-Scorevalues (See Fig. 5).

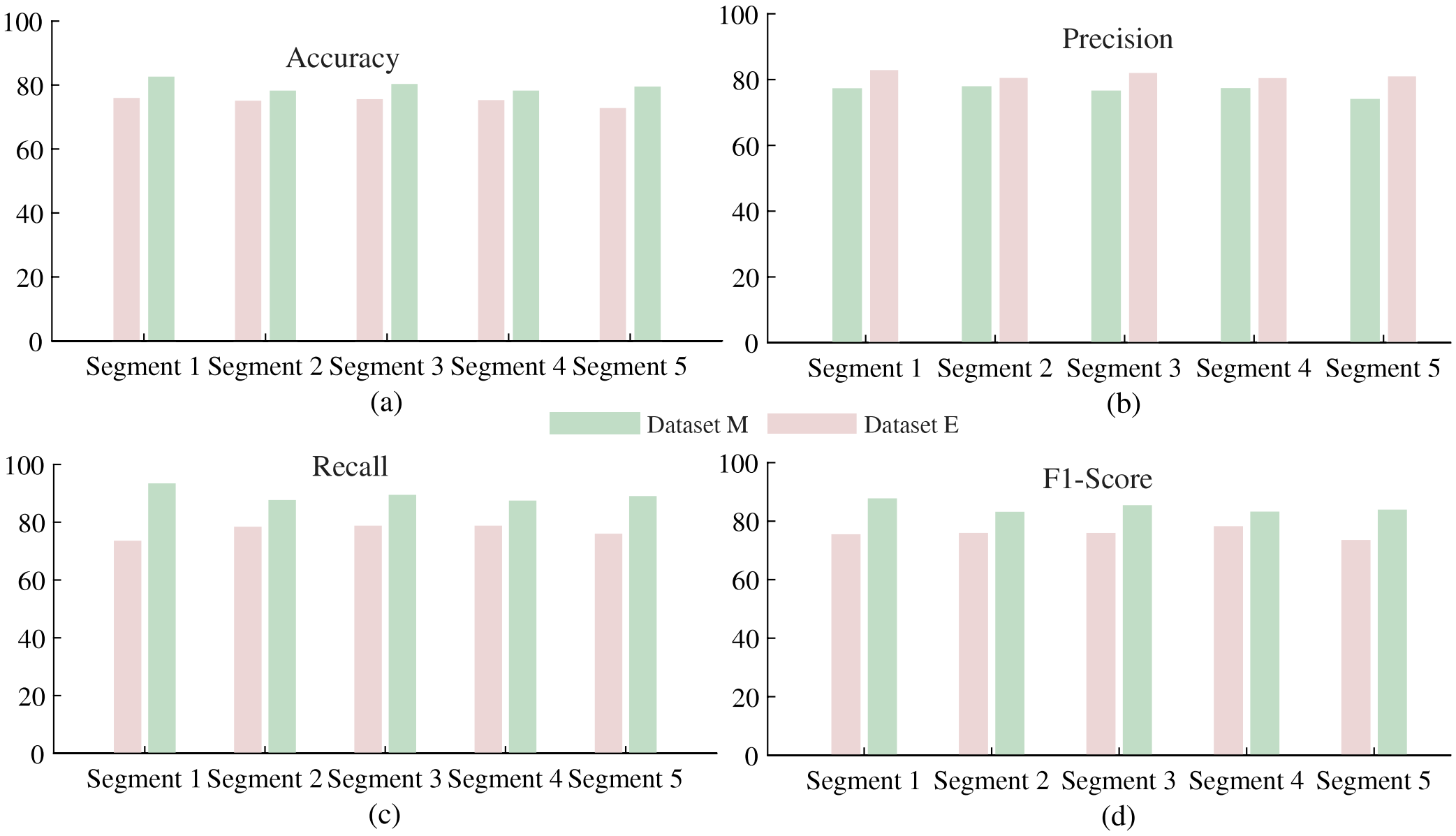

**Fig. 3.** The results of LOSO-CV extracted from dataset M and dataset across five consecutive 1-minute gameplay segments. (a) Accuracy (b) Precision (c) Recall (d) F1-Score.

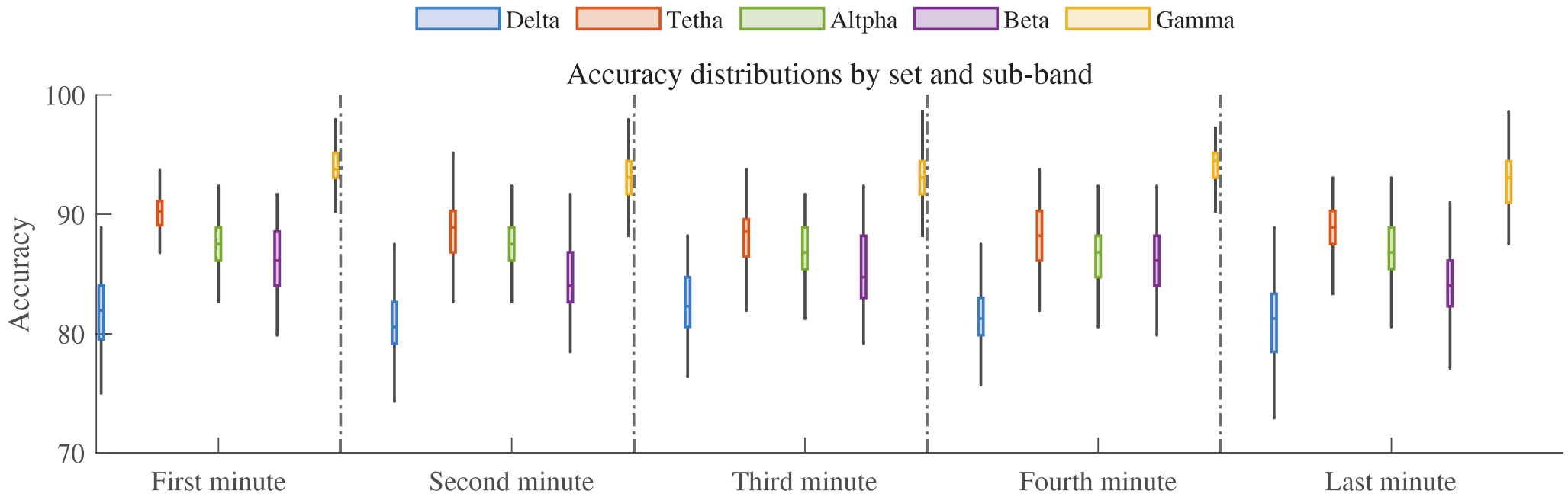

**Fig. 4.** Dataset E: boxplots of accuracy across EEG sub-bands ($\delta$, $\theta$, $\alpha$, $\beta$, $\gamma$) for the five 1-min segments, showing consistent band ranking and tight dispersion.

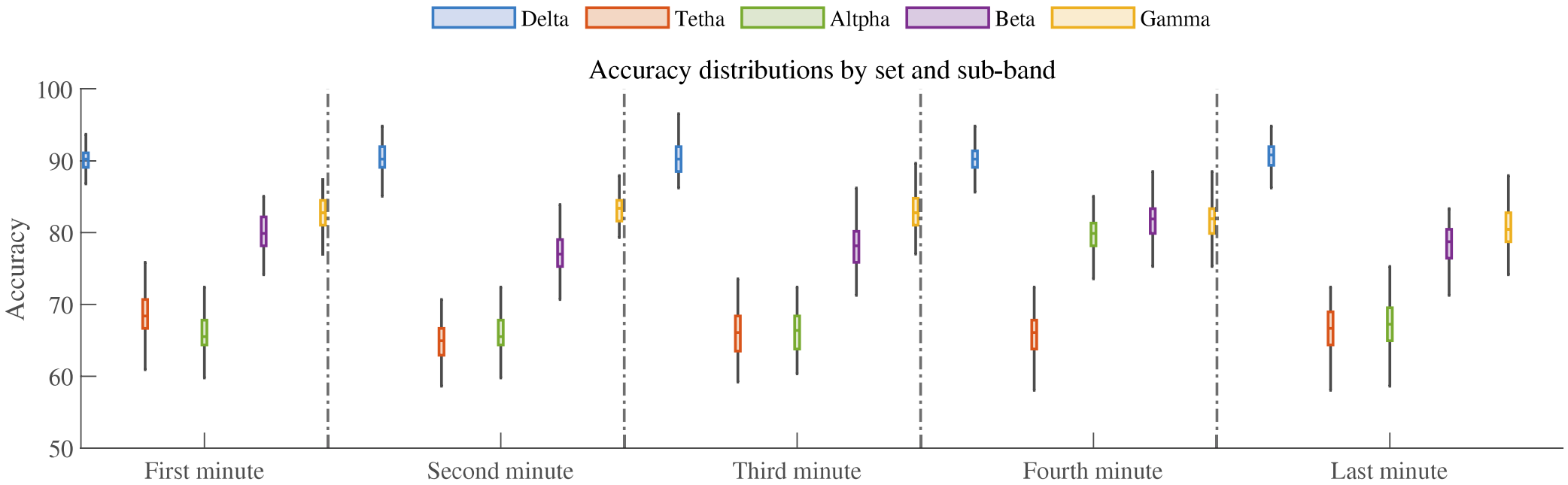

**Fig. 5.** Dataset M: boxplots of accuracy across EEG sub-bands ($\delta$, $\theta$, $\alpha$, $\beta$, $\gamma$) for the five 1-min segments, showing consistent band ranking and tight dispersion.

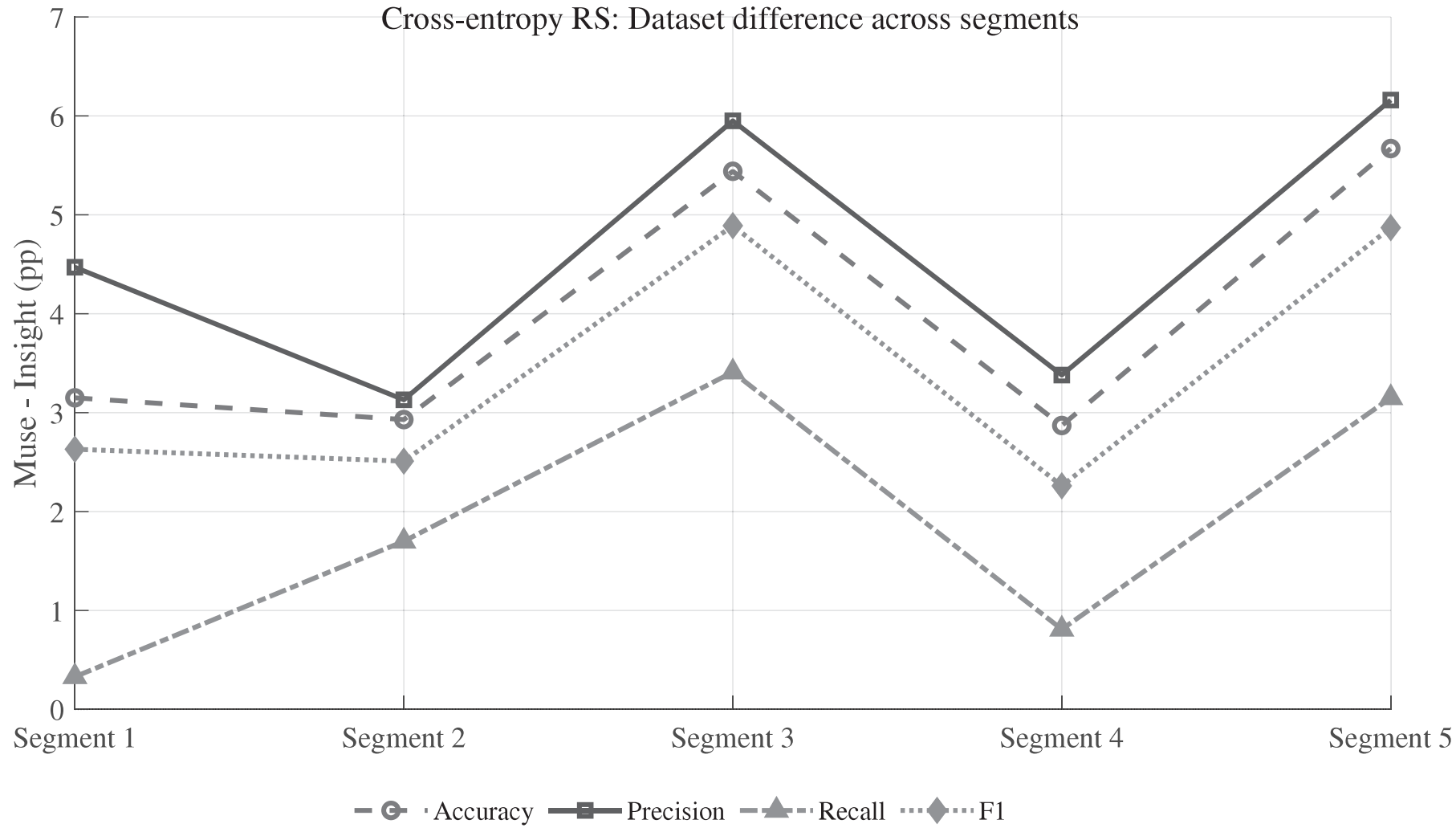

**Fig. 6.** Segment-wise comparison of Accuracy, Precision, Recall, and F1-score under random sampling: Dataset M consistently exceeds Dataset E across all segments.

Motivated by the strong performance of the Random Forest baseline, we evaluated six additional machine learning classifiers on both datasets. The results for all six methods are presented and compared in Fig. 7. The experimental results for Dataset E demonstrate that the SVM with RBF kernel consistently achieved the highest accuracy across segments, ranging from 83.70% to 87.21%, followed closely by GentleBoost with accuracy between 82.98% and 85.72%. Precision values for these classifiers also remained high, with SVM-RBF precision spanning 81.80% to 85.07%, and GentleBoost ranging from 81.59% to 83.96%. Recall scores revealed a similar pattern, with SVM-RBF achieving recall rates up to 92.35% and GentleBoost up to 89.48%, indicating strong sensitivity. F1-scores for these methods were robust as well, peaking at 88.36% for SVM-RBF and maintaining solid values for GentleBoost. While other classifiers like SVM Linear, fitted discriminant, kNN, and Naive Bayes showed somewhat lower but still respectable performances, accuracy mostly between 57.93% and 75.52%, all models demonstrated reasonably consistent results. Although these metrics do not surpass those typically achieved with Random Forest approaches, the overall consistency and strength across multiple machine learning algorithms indicate that the identified channel-wise entropies are reliable and stable predictors, applicable across different modeling frameworks and data segments (Fig. 7). Across both datasets, the overall pattern is consistent: SVM-based classifiers (RBF and linear) and boosting methods sit in the top tier across segments, kNN and fitted discriminant models occupy the middle, and the naive baseline trails on all four metrics (accuracy, precision, recall, F1). Performance is fairly stable across the five 1-min segments, with mid-session windows (Segments 2–4) tending to be the strongest. In both datasets, recall generally exceeds precision, so F1-Score falls between them, mirroring the usual relationship among these metrics. Taken together, the two sets of plots tell a similar story about model ranking and temporal stability, strongest results for SVM/boosting, weakest for the Naive Bayes classifier, and modest segment-to-segment variation concentrated in the middle of the session (Fig. 7).

## 5. Discussions and conclusions

Flow is a psychological state that arises during challenging activities when the task demands are well matched to an individual's skill level [26]. This work demonstrates robust discrimination between flow and rest periods, which is feasible using consumer-grade, low-channel EEG and entropy-based biomarkers. Using two consumer-oriented EEG headsets (Emotiv Insight, Muse-S), wavelet denoising, sub-band decomposition ($\delta$, $\theta$, $\alpha$, $\beta$, $\gamma$), and a compact feature set (slope, spectral, distribution, and cross-channel entropies), we obtained mean accuracies up to 97% with RS-CV and 82.75% with LOSO-CV, the latter providing a conservative estimate of out-of-subject generalization. Across datasets, "all-entropy" models were consistently strongest; within single type features, SpecEn and DisEn outperformed SlopEn, and the mid-session segments (2–4) tended to yield the best results, indicating stable discriminative information once gameplay is underway. Beyond Random Forest baselines, SVMs and boosting methods remained the top non-RF alternatives, while kNN and fitted discriminant were mid-tier and Naive Bayes models lagged, reinforcing that the learned representations transfer across algorithmic families. Together, these findings show that simple frontal montages plus information-theoretic features can support reliable, headset-agnostic flow detection in realistic settings.

Comparing to state-of-the-art research regarding the neural correlates of flow, particularly the role of transient hypofrontality characterized by reduced prefrontal cortex activity that dampens analytical self-reflection while enhancing procedural, skill-based processing, this study advances our understanding of flow mechanisms [16]. While previous fMRI studies have shown flow-related activation patterns in the anterior insula, inferior frontal gyri, and basal ganglia with concurrent deactivation in medial prefrontal regions [17], the present work demonstrates that these neurological signatures can be captured through accessible, easy to use EEG devices. This is particularly significant given that earlier EEG-based flow detection studies employing full-cap, multi-channel configurations such as those documenting correlations between flow scores and delta, theta, and gamma power, achieved promising results but at the cost of practical deployment in everyday settings [20]. By contrast, the current approach using only two to five prefrontal channels from consumer headsets addresses the critical gap between laboratory-based flow research and real-life applicability.

The superiority of entropy-based features over traditional spectral power measures represents a methodological contribution that differentiates this work from existing literature. Entropy measures characterize signal complexity and irregularity rather than neural activation magnitude. Consequently, the present findings do not provide direct evidence for prefrontal hypoactivation. However, altered entropy patterns observed over prefrontal electrodes are consistent with changes in neural organization expected under transient hypofrontality and may reflect functional reconfiguration of prefrontal processing during flow. While studies by [20,21] successfully used spectral power features to correlate with subjective flow reports, the present investigation demonstrates that combining channel-wise entropy measures specifically Slope Entropy, Distribution Entropy, and Spectral Entropy with cross-channel entropy features yields substantially higher classification accuracy. The cross-channel entropy measures, which quantify functional coupling and network-level dynamics between electrode pairs and frequency sub-bands, provided complementary information regarding inter-channel interactions during flow. Although their standalone classification performance was lower than that achieved using the complete channel-wise entropy feature set, they consistently demonstrated discriminative capability across both datasets and may offer additional insights into network-level neural organization associated with flow states. This finding complements research by Rácz et al. [15], who demonstrated that multimodal biosignals (EEG, HRV, GSR, and motion tracking) provide complementary flow-state information, although the current study achieves comparable discrimination using EEG entropy features alone.

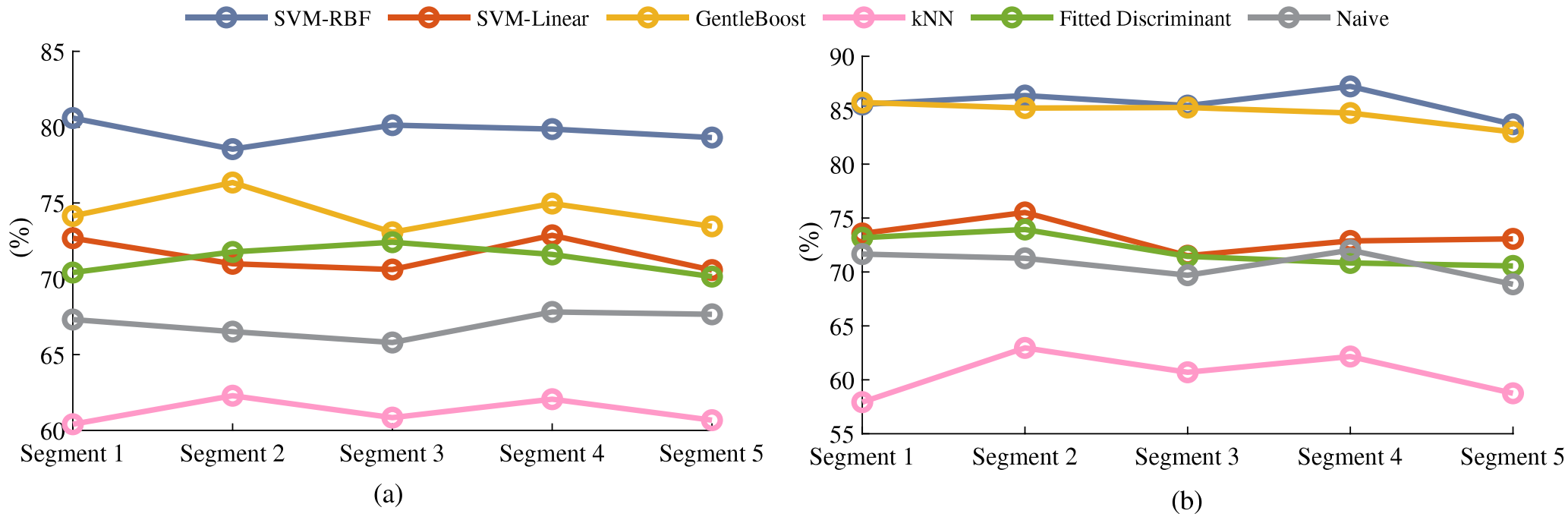

**Fig. 7.** Performance of six classifiers: SVM-RBF, SVM-Linear, GentleBoost, kNN, Fitted Discriminant, and a Naive baseline across five 1-min segments. Panels show (a) Accuracy of each machine learning method across dataset M, (b)Accuracy of each classifier across dataset E. Overall, SVMs and boosting consistently achieve the highest scores, kNN and fitted discriminant are midtier, and the naïve model performs worst. Performance is relatively stable over time, with mid-session segments (2–4) generally strongest and recall exceeding precision, yielding F1-Scorevalues between them.

The temporal analysis revealing peak performance during mid-session segments (minutes 2–4 of gameplay) offers novel insight into the dynamics of flow state emergence and maintenance. This pattern suggests that flow requires a brief warm-up period and may be susceptible to fatigue effects toward the end of extended engagement, a finding that warrants further investigation and has implications for optimal task design. Furthermore, the consistent frequency-band ranking observed across datasets (gamma and theta bands showing the strongest discriminative power in the dataset E; delta band dominant in the dataset M) corroborates earlier reports linking increased delta and theta power to flow states while extending this knowledge to consumer-grade devices with varying electrode configurations. The comprehensive validation across multiple machine learning architectures, Random Forest, SVM with RBF and linear kernels, GentleBoost, k-NN, fitted discriminant classifiers, and Naive Bayes demonstrates the robustness and generalizability of the extracted entropy features. The consistent performance hierarchy across classifiers (SVM-RBF and GentleBoost achieving 83%–87% accuracy, substantially exceeding naive baselines) provides strong evidence that the identified biomarkers capture genuine neurophysiological signatures rather than dataset-specific artifacts. This multi-classifier validation addresses a limitation noted in prior wearable EEG flow studies [20,22,24] which often relied on single classification approaches with modest predictive power and limited generalizability. The superior performance of Dataset M (Muse) compared to Dataset E (Emotiv Insight) in all validation schemes — with consistent advantages of 2–5 percentage points, probably reflects differences in signal quality, electrode positioning and sampling rates between the two devices. An additional factor may be the difference in channel count. The Emotiv headset contains five channels, increasing both spatial information and susceptibility to inter-subject variability. Future work should evaluate matched-channel analyses to isolate the influence of electrode count from other device-specific factors. The Muse headset's focus on prefrontal regions (AF7, AF8) may provide more stable signals from the DLPFC and MPFC, brain regions critically implicated in flow experiences. Conversely, the Emotiv's broader spatial coverage, while potentially informative, may introduce greater inter-subject variability in electrode placement and impedance. The marked difference between RS-CV and LOSO-CV results highlights the substantial inter-individual variability in flow-related EEG signatures, a finding consistent with research on cardiovascular and respiratory flow biomarkers by [5,9], who similarly observed higher classification accuracy in naturalistic field studies compared to controlled laboratory settings. This variability likely stems from anatomical differences, cognitive strategies, and individual flow susceptibility. The LOSO-CV results, while more conservative, still represent a substantial improvement over chance and demonstrate genuine cross-subject generalizability a critical requirement for practical deployment that was insufficiently addressed in earlier wearable EEG flow studies.

Despite the encouraging results, several limitations merit acknowledgment. First, the reliance on Tetris gameplay as the flow-induction paradigm, while well-established in the literature, limits generalizability to other flow-inducing activities that engage different cognitive, emotional, and sensorimotor capacities. Flow during athletic performance, creative tasks, or professional work may exhibit distinct neural signatures requiring validation. Second, the dependence on post-trial subjective flow ratings introduces potential reporting biases and precludes real-time flow monitoring—a capability essential for applications such as adaptive difficulty adjustment in educational software or performance optimization in training environments. Finally, the relatively modest sample sizes (29 participants for Muse; 16 for Emotiv) constrain statistical power for detecting subtle individual differences and evaluating rare flow subtypes. Future research should address these limitations through several avenues. Expanding the range of flow-induction tasks to include diverse domains (sports, music, creative work, learning) would establish the cross-domain validity of entropy-based biomarkers. Developing real-time flow detection algorithms with minimal computational latency would enable closed-loop applications, such as adaptive training systems that dynamically adjust task difficulty to maintain optimal flow conditions—an approach anticipated by the transient hypofrontality framework. Larger, more demographically diverse samples would permit investigation of age, gender, and expertise-related differences in flow signatures, potentially enabling personalized flow detection models. From a practical standpoint, this work demonstrates that consumer EEG technology has matured to a level where meaningful mental state detection is achievable outside controlled laboratory settings, a development with far-reaching implications for education, workplace productivity, sports training, and mental health interventions. In conclusion, this study makes three principal contributions to the field of flow state detection. First, it validates entropy-based EEG features as robust, device-independent biomarkers of flow that outperform traditional spectral power measures. Second, it demonstrates that consumer-grade prefrontal EEG headsets can achieve clinically meaningful classification accuracy, bridging the gap between laboratory research and real-world deployment. Third, it establishes methodological best practices for wearable EEG flow detection, including wavelet-based artifact removal, multi-entropy feature extraction, and rigorous subject-independent validation. These findings advance the neurophysiological understanding of flow states while providing a practical foundation for next-generation flow monitoring technologies that can enhance human performance, learning, and well-being across diverse domains of activity. Lastly, although wavelet-based hard-threshold denoising is widely used for suppressing ocular artifacts in EEG recordings, it may also alter signal complexity characteristics that contribute to entropy estimation. Because the present study did not explicitly quantify the effect of denoising on the extracted entropy features, it cannot be excluded that some feature distortion occurred. Future work should systematically evaluate the robustness of entropy-based flow biomarkers under different artifact-removal strategies and compare feature stability before and after preprocessing.

## CRediT authorship contribution statement

**Matin Beiramvand:** Writing – original draft, Visualization, Project administration, Methodology, Investigation, Funding acquisition, Formal analysis, Data curation, Conceptualization. **Reijo Koivula:** Data curation, Conceptualization. **Tarmo Lipping:** Supervision, Methodology, Formal analysis, Data curation.

## Declaration of competing interest

The authors declare the following financial interests/personal relationships which may be considered as potential competing interests: Matin Beiram Vand reports financial support was provided by Tampere University and Tampere University of Applied Sciences. If there are other authors, they declare that they have no known competing financial interests or personal relationships that could have appeared to influence the work reported in this paper.

## Data availability

The authors do not have permission to share data.

## References

[1] M. Csikszentmihalyi, Beyond Boredom and Anxiety: Experiencing Flow in Work and Play, Jossey-Bass, San Francisco, 1975.

[2] J. Macbeth, Ocean cruising,optimal experience: Psychological studies of flow in consciousness, Camb. Univ. Press; Camb. (UK) (1988) 214–231, doi:Ocean-cruising/991005543143707891.

[3] I. Sato, Bosozoku: flow in Japanese motorcycle gangs, in: M. Csikszentmihalyi, I.S. Csikszentmihalyi (Eds.), Optimal Experience: Psychological Studies of Flow in Consciousness, Cambridge University Press, 1988, pp. 92–117.

[4] S.A. Jackson, S.K. Ford, J.C. Kimiecik, H.W. Marsh, Psychological correlates of flow in sport, "J. Sport. Exerc. Psychol." 20 (4) (1998) 358–378, http://dx.doi.org/10.1123/jsep.20.4.358, URL https://journals.humankinetics.com/view/journals/jsep/20/4/article-p358.xml.

[5] J. Keller, H. Bless, Flow and regulatory compatibility: An experimental approach to the flow model of intrinsic motivation, Pers. Soc. Psychol. Bull. 34 (2) (2008) 196–209, http://dx.doi.org/10.1177/0146167207310026, PMID: 18212330. arXiv:https://doi.org/10.1177/0146167207310026.

[6] J. Keller, H. Bless, F. Blomann, D. Kleinböhl, Physiological aspects of flow experiences: Skills-demand-compatibility effects on heart rate variability and salivary cortisol, J. Exp. Soc. Psychol. 47 (4) (2011) 849–852, http://dx.doi.org/10.1016/j.jesp.2011.02.004, URL https://www.sciencedirect.com/science/article/pii/S0022103111000321.

[7] Ö. de Manzano, T. Theorell, L. Harmat, F. Ullén, The psychophysiology of flow during piano playing, Emotion 10 (3) (2010) 301–311, http://dx.doi.org/10.1037/a0018432.

[8] M.T. Allison, M.C. Duncan, Women, work, and flow, in: M. Csikszentmihalyi, I.S. Csikszentmihalyi (Eds.), Optimal Experience: Psychological Studies of Flow in Consciousness, Cambridge University Press, Cambridge, UK and New York, NY, 1988, pp. 118–137, http://dx.doi.org/10.1017/CBO9780511621956.007.

[9] R. Rissler, M. Nadj, M.X. Li, N. Loewe, M.T. Knierim, A. Maedche, To be or not to be in flow at work: Physiological classification of flow using machine learning, IEEE Trans. Affect. Comput. 14 (1) (2023) 463–474, http://dx.doi.org/10.1109/TAFFC.2020.3045269.

[10] C. Besson, D. Aubert, Z. Lavorato, R. Billieux, J.L. Clénin, R. Donzè, P. Schmitt, F. Deriaz, H. Luthi, T. Decosterd, P.G. Zimmermann, Assessing the clinical reliability of short-term heart rate variability: insights from controlled dual-environment and dual-position measurements, Sci. Rep. 15 (1) (2025) 89892, http://dx.doi.org/10.1038/s41598-025-89892-3, URL https://www.nature.com/articles/s41598-025-89892-3.

[11] C. Alameda, D. Sanabria, L.F. Ciria, The brain in flow: A systematic review on the neural basis of the flow state, Cortex 154 (2022) 348–364, http://dx.doi.org/10.1016/j.cortex.2022.06.005, URL https://www.sciencedirect.com/science/article/pii/S0010945222001836.

[12] S. Jha, N. Stogios, A.S. de Oliveira, S. Thomas, R.P. Nolan, Getting into the zone: A pilot study of autonomic-cardiac modulation and flow state during piano performance, Front. Psychiatry 13 (2022) 853733, http://dx.doi.org/10.3389/fpsyt.2022.853733, URL https://pmc.ncbi.nlm.nih.gov/articles/PMC9044034/.

[13] S. Khoshnoud, F. Alvarez Igarzábal, M. Wittmann, Brain-heart interaction and the experience of flow while playing a video game, Front. Hum. Neurosci. 16 (2022) 819834, http://dx.doi.org/10.3389/fnhum.2022.819834, URL https://pmc.ncbi.nlm.nih.gov/articles/PMC9096496/.

[14] E. Gaston, F. Ullén, L.W. Wesseldijk, M.A. Mosing, Can flow proneness be protective against mental and cardiovascular health problems? A genetically informed prospective cohort study, Transl. Psychiatry 14 (1) (2024) 144, http://dx.doi.org/10.1038/s41398-024-02855-6, URL https://pmc.ncbi.nlm.nih.gov/articles/PMC10937942/.

[15] M. Rácz, M. Becske, T. Magyaródi, G. Kitta, M. Szuromi, G. Márton, Physiological assessment of the psychological flow state using wearable devices, Sci. Rep. 15 (1) (2025) 11839, http://dx.doi.org/10.1038/s41598-025-95647-x, URL https://pmc.ncbi.nlm.nih.gov/articles/PMC11977251/.

[16] A. Dietrich, Neurocognitive mechanisms underlying the experience of flow, Conscious. Cogn. 13 (4) (2004) 746–761, http://dx.doi.org/10.1016/j.concog.2004.07.002, URL https://www.sciencedirect.com/science/article/pii/S1053810004000583.

[17] M. Ulrich, J. Keller, G. Grön, Neural signatures of experimentally induced flow experiences identified in a typical fMRI block design with BOLD imaging, Soc. Cogn. Affect. Neurosci. 11 (3) (2015) 496–507, http://dx.doi.org/10.1093/scan/nsv133, arXiv:https://academic.oup.com/scan/article-pdf/11/3/496/27103194/nsv133.pdf.

[18] M. Ulrich, J. Keller, K. Hoenig, C. Waller, G. Grön, Neural correlates of experimentally induced flow experiences, NeuroImage 86 (2014) 194–202, http://dx.doi.org/10.1016/j.neuroimage.2013.08.019, URL https://www.sciencedirect.com/science/article/pii/S1053811913008732.

[19] M.D. Ferrell, R.L. Beach, N.M. Szeverenyi, M. Krch, B. Fernhall, An fMRI analysis of neural activity during perceived zone-state performance, "J. Sport. Exerc. Psychol." 28 (4) (2006) 421–433, http://dx.doi.org/10.1123/jsep.28.4.421, URL https://journals.humankinetics.com/view/journals/jsep/28/4/article-p421.xml.

[20] Y. Hang, B. Unenbat, S. Tang, F. Wang, B. Lin, D. Zhang, Exploring the neural correlates of flow experience with multifaceted tasks and a single-channel prefrontal EEG recording, Sensors 24 (6) (2024) 1894, http://dx.doi.org/10.3390/s24061894, URL https://pmc.ncbi.nlm.nih.gov/articles/PMC10975495/.

[21] A. Leroy, G. Cheron, EEG dynamics and neural generators of psychological flow during one tightrope performance, Sci. Rep. 10 (2020) http://dx.doi.org/10.1038/s41598-020-69448-3, URL https://www.nature.com/articles/s41598-020-69448-3.

[22] M.T. Irshad, F. Li, M.A. Nisar, X. Huang, M. Buss, L. Kloep, C. Peifer, B. Kozusznik, A. Pollak, A. Pyszka, O. Flak, M. Grzegorzek, Wearable-based human flow experience recognition enhanced by transfer learning methods using emotion data, Comput. Biol. Med. 166 (2023) 107489, http://dx.doi.org/10.1016/j.compbiomed.2023.107489, URL https://www.sciencedirect.com/science/article/pii/S001048252300954X.

[23] M. Çınaroğlu, Hormonal catalysts in the addiction cycle of muscle dysmorphia: A neuroendocrine perspective, J. Neurobehav. Sci. 11 (1) (2024) 1–9, http://dx.doi.org/10.4103/jnbs.jnbs_19_23.

[24] G. Rosso, R. Ricci, L. Pia, G. Rebaudo, M. Guindani, A. Marocchino, G. De Pieri, A.F. Rosso, Quantifying flow state dynamics: A prefrontal cortex EEG-based model validation study. Unveiling the prefrontal cortex's role in flow state experience: An empirical EEG analysis, 2025, http://dx.doi.org/10.48550/arXiv.2506.16838, ArXiv. arXiv:2506.16838.

[25] E. Sajno, Affective computing for detecting psychological flow state: a definition and methodological problem, in: 2023 11th International Conference on Affective Computing and Intelligent Interaction Workshops and Demos, ACIIW, 2023, pp. 1–5, URL https://api.semanticscholar.org/CorpusID:267023097.

[26] L. Harmat, Ö. de Manzano, T. Theorell, L. Högman, H. Fischer, F. Ullén, Physiological correlates of the flow experience during computer game playing, Int. J. Psychophysiol. 97 (1) (2015) 1–7, http://dx.doi.org/10.1016/j.ijpsycho.2015.05.001, URL https://www.sciencedirect.com/science/article/pii/S0167876015001683.

[27] M. Beiramvand, R. Koivula, T. Lipping, Development of an EEG-based method for detecting flow state using a wearable headband in a game environment*, in: 2025 47th Annual International Conference of the IEEE Engineering in Medicine and Biology Society, EMBC, 2025, pp. 1–6, http://dx.doi.org/10.1109/EMBC58623.2025.11251885.

[28] Muse, https://https://choosemuse.com/.

[29] Insight, https://www.emotiv.com/products/insight.

[30] Petal, https://petal.tech/.

[31] M. Beiramvand, M. Shahbakhti, N. Karttunen, R. Koivula, J. Turunen, T. Lipping, Assessment of mental workload using a transformer network and two prefrontal EEG channels: An unparameterized approach, IEEE Trans. Instrum. Meas. 73 (2024) 1–10, http://dx.doi.org/10.1109/TIM.2024.3395312.

[32] M. Shahbakhti, M. Beiramvand, S.M. Far, J. Solé-Casals, T. Lipping, P. Augustyniak, Utilizing slope entropy as an effective index for wearable EEG-based depth of anesthesia monitoring, in: 2024 46th Annual International Conference of the IEEE Engineering in Medicine and Biology Society, EMBC, 2024, pp. 1–4, http://dx.doi.org/10.1109/EMBC53108.2024.10782706.

[33] E. Gani, N. Handayani, S. Harke Pratama, N. Faadhilah Afif, F. Aziezah, A. Christy Keintjem, F. Haryanto, Suprijadi, Brainwaves analysis using spectral entropy in children with autism spectrum disorders (ASD), J. Phys.: Conf. Ser. 1505 (1) (2020-03-01) 012070, http://dx.doi.org/10.1088/1742-6596/1505/1/012070, URL https://iopscience.iop.org/article/10.1088/1742-6596/1505/1/012070.

[34] P. Li, C. Karmakar, C. Yan, M. Palaniswami, C. Liu, Classification of 5-s epileptic EEG recordings using distribution entropy and sample entropy, Front. Physiol. 7 (2016) 136, http://dx.doi.org/10.3389/fphys.2016.00136, URL https://pmc.ncbi.nlm.nih.gov/articles/PMC4830849/.

[35] J.F. Valencia, A. Porta, M. Vallverdu, F. Claria, R. Baranowski, E. Orlowska-Baranowska, P. Caminal, Refined multiscale entropy: Application to 24-h holter recordings of heart period variability in healthy and aortic stenosis subjects, IEEE Trans. Biomed. Eng. 56 (9) (2009) 2202–2213, http://dx.doi.org/10.1109/TBME.2009.2021986.

[36] T.M. Cover, J.A. Thomas, Elements of information theory, Wiley-Interscience (1991).

[37] S. Khatun, B.I. Morshed, G.M. Bidelman, A single-channel EEG-based approach to detect mild cognitive impairment via speech-evoked brain responses, IEEE Trans. Neural Syst. Rehabil. Eng. 27 (5) (2019) 1063–1070, http://dx.doi.org/10.1109/TNSRE.2019.2911970.

[38] M. Zhou, H. Zhang, W. Zhang, Y. Yi, An improved random forest algorithm-based fatigue recognition with multiphysical feature, IEEE Sensors J. 23 (21) (2023) 26195–26201, http://dx.doi.org/10.1109/JSEN.2023.3314316.

[39] Y. Ma, X. Ding, Q. She, Z. Luo, T. Potter, Y. Zhang, Classification of motor imagery EEG signals with support vector machines and particle swarm optimization, Comput. Math. Methods Med. 2016 (2016) 4941235, http://dx.doi.org/10.1155/2016/4941235.

[40] J.H. Friedman, T. Hastie, R. Tibshirani, Additive logistic regression: a statistical view of boosting, Ann. Stat. 28 (2) (2000) 337–407, http://dx.doi.org/10.1214/aos/1016218223.

[41] S. Wu, et al., Adaptive LDA classifier enhances real-time control of an imagined speech brain-computer interface, IEEE Trans. Neural Syst. Rehabil. Eng. 32 (2024) 1234–1243, http://dx.doi.org/10.1109/TNSRE.2024.3356789.

[42] M.-P. Hosseini, A. Hosseini, K. Ahi, A review on machine learning for EEG signal processing in bioengineering, IEEE Rev. Biomed. Eng. 14 (2021) 204–218, http://dx.doi.org/10.1109/RBME.2020.2969915.